# Anti-retroviral therapy for HIV: who should we test and who should we treat?†

Brian G. Williams, Renier van Rooyen and Martin Nieuwoudt

South African Centre for Epidemiological Modelling and Analysis, Stellenbosch, South Africa

Correspondence to BrianGerardWilliams@gmail.com

**Abstract**

In most countries CD4$^+$ cell counts are still used for deciding when to start HIV-positive people on anti-retroviral therapy. However, various CD4$^+$ thresholds, 200, 350 or 500/μL, are chosen arbitrarily and for historical reasons. Here we consider the optimal CD4$^+$ threshold at which asymptomatic HIV-positive people living in Botswana, South Africa and Zimbabwe should start treatment depending on their prognosis given their CD4$^+$ cell counts or viral load. We also examine the optimal interval at which people should be retested if they are HIV-negative. This analysis shows that while the use of CD4$^+$ cell counts or viral load tests could have been useful in deciding how to triage patients for treatment at the start of the epidemic this is no longer the case except possibly for those aged about 15 to 25 years. In order not to do harm to individual patients everyone should be started on ART as soon as they are found to be HIV-positive.

## Introduction

Deciding when to start anti-retroviral therapy is important both for the prognosis of the individual patient and to prevent transmission. In this paper we are concerned only with the prognosis for asymptomatic, HIV-positive people (WHO clinical stage I[1,2]); we assume that all those with AIDS defining infections or conditions, (WHO clinical stages II to IV[1,2]) should start anti-retroviral therapy (ART) immediately.

In 1996 triple-drug therapy became available[3,4] and was shown to reduce plasma viral load by two orders of magnitude[5] but the annual cost per person for first-line triple-drug therapy in Africa was US$12k[6] and some of the drugs gave rise to serious side effects. Nevertheless, in 1996 the International AIDS Society (IAS) recommended giving ART to all patients with a CD4$^+$ cell count below 500/μL or a viral load greater than 30k/mL.[4] In 1997 the IAS dropped the CD4$^+$ condition but lowered the viral load threshold to 5k/mL. They noted that a decision to initiate therapy at CD4$^+$ cell counts above 500/μL should be tempered by the potential problems related to long-term toxicity, tolerance, acceptance, cost, and the development of drug resistance.[4] Nevertheless, it was clear that early initiation of anti-retroviral therapy (ART) had virological, immunological and clinical benefits[4] and the IAS maintained their recommendations for starting ART, with slight variations, until the year 2000.[7-9] In the year 2000 the IAS reintroduced the CD4$^+$ threshold at 350/μL and maintained the viral load cut-off at 5k/mL.

In 2002 there was an important shift towards delaying treatment because some key drugs were poorly tolerated leading the IAS to lower the CD4$^+$ threshold and raise the viral load threshold. ART was now recommended for those with a CD4$^+$ cell count less than 200/μL or a viral load greater than 50k/mL to 100k/mL.[10]

At about the same time the World Health Organization (WHO), in collaboration with the United States National Institutes of Health, convened consultative meetings in which more than two hundred clinicians, scientists, government representatives, representatives of civil society and people living with HIV/AIDS from more than sixty countries participated.[11] The recommendations noted that while starting therapy at CD4 cell counts above 200/μL clearly provides clinical benefits, the actual point above 200/μL at which to start therapy had not been definitively determined.[11] For reasons that are not made explicit in the published guidelines[11] WHO recommended that treatment should be given only to people with a CD4$^+$ cell count below 200/μL, in agreement with the IAS recommendations. Because viral load testing was not generally available in resource limited settings, the viral load condition was not included. In 2003 WHO revised their guidelines slightly upwards and suggested using CD4 cell counts between 200/μL and 350/μL to assist decision-making for those in WHO clinical stages III or IV.[12] In 2007 WHO published a further revision[13] in which they said that at a CD4$^+$ cell count below 200/μL treatment should be started immediately, between 200/μL and 350/μL treatment should be considered while taking into account clinical staging, and that above 350/μL treatment should not be started.

Following a review and synthesis of new evidence the World Health Organization revised their guidelines in 2009 to recommend starting people on ART if their CD4$^+$ cell count is less than 350/μL.[14,15]

In 2010 the IAS once again recommended ART for all HIV-positive people with a CD4$^+$ cell count below 500/μL, without excluding treatment for those with a higher cell count, or a viral load above 100k/mL.[16] In 2012 the IAS recommended ART for all HIV-positive people.[17]

By 2001 the annual cost of triple drug therapy had dropped form US$12k to less than US$1k and the price of drugs has continued to fall[6,18] with generic formulations available in 2012 for US$150 per year.[18]

---

† This is an expanded version of a presentation given by Brian Williams at the 2nd International HIV Treatment as Prevention (TasP) Workshop, April 22–25, 2012, Vancouver, BC, Canada

The new drugs are easier to tolerate and have few serious side effects.

In order to make the implications of the various changes in recommendations clear we use $CD4^+$ and plasma viral load data from a cross-sectional study of young men in Orange Farm, South Africa, in 2002[19,20] to estimate the proportion that would have been eligible for treatment under the different, changing, recommendations. The results are shown in Figure 1.

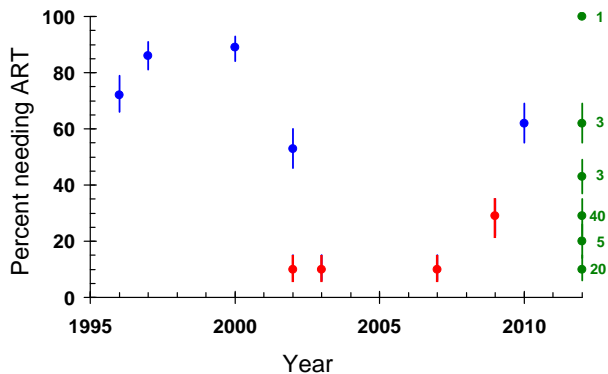

Figure 1. The proportion of young men in Orange Farm, South Africa, who would had to have started treatment in 2002 as the treatment guidelines changed over time. Blue: IAS; red: WHO; green current practice in different numbers of countries. Error bars reflect the sample size in the Orange Farm data.

Using the 19 96 IAS guidelines[4] ($CD4^+$ cut-off: 500/μL; viral load cut-off: 30k/mL), 72% (66%−79%) of all HIV-positive men in Orange Farm in 2002 should have started ART. Under the 1997 IAS guidelines[8] (viral load cut-off: 5k/mL), the proportion needing to start treatment would have increased to 86% (81%−90%) and under the 2000 IAS guidelines[6] ($CD4^+$ cut-off: 500/μL; viral load cut-off: 5k/mL) would have increased further to 89% (84%−93%). The 2002 IAS guidelines[10] ($CD4^+$ cut-off: 200/μL; viral load cut-off: 50k/mL) were more conservative and the proportion needing to start treatment would have fallen to 53% (46%−60%). The 2002 WHO guidelines were considerably more conservative[11] ($CD4^+$ cut-off: 200/μL) because they did not include a viral load criterion and the proportion of people eligible for ART would have dropped dramatically to only 10% (6%−15%) of those living with HIV. The 2003[12] and 2007[13] WHO guidelines remained the same for asymptomatic people while the 2009 WHO guidelines[14,15] were a little more aggressive ($CD4^+$ cut-off: 350/μL) and the proportion needing treatment would have increased to 29% (22%−35%). The 2010 IAS guidelines[16] ($CD4^+$ cut-off: 500/ml; viral load cut-off 100k/mL) would have increased this further to 62% (55%−69%) and finally the 2012 IAS guidelines[17] recommend immediate treatment for all HIV-positive people.

In practice the guidelines for when to start treatment now vary greatly by country as indicated by the green dots in Figure 1.[21] In the United States of America the advice is to start ART immediately, irrespective of $CD4^+$ cell count. Three countries recommend starting ART at a $CD4^+$ cell count at or below 500/μL, 40 recommend starting ART at a $CD4^+$ cell count at or below 350/μL and in 3 a threshold of 500/μL may also be considered. In 5 countries ART initiation is only recommended for those with a $CD4^+$ cell count below 250/μL and in 20 countries below 200/μL.[21]

While several studies have shown that the earlier HIV-positive people start treatment the better is their prognosis[22-30] this has not been fully examined in relation to the $CD4^+$ cell count or the viral load at which people start ART. There is still a need to provide a rational basis for deciding on when people should start ART if their $CD4^+$ cell counts or viral load are known and for deciding how often people should return to be tested if they are HIV-negative. In this paper we consider the prognostic value of $CD4^+$ cell counts and plasma viral load in relation to the likely survival of a person infected with HIV in three African countries.

ART can reduce the viral load in people infected with HIV by up to 10,000 times,[31-33] and this can be sustained for at least seven years[31,32] and probably indefinitely. It therefore follows that annual testing and immediate ART can stop transmission of HIV.[33-38] However, it is important to ensure that public health benefits to the wider community are not gained at the expense of significant harm to individual patients. Here we do not consider the broader public health benefits that may accrue from early treatment of HIV-positive people[34,36,39] but focus on what is in the best interests of the individual patient.

## Methods

It is important for countries to decide on sensible criteria for starting people on ART. Here we suggest that this should be based on the assumption that where resources are limited we need to prioritise those people that are most likely to die soon. In an ideal world one would simply rank people in order of their predicted life expectancy and then start from those with the shortest life-expectancy and progressively include people with longer and longer life expectancies until the supply of drugs is exhausted. In practice we cannot do this because we do not have complete information on all HIV-positive people at a given time and indeed some people will be found to have HIV while they are still healthy while others will only be found to have HIV when they are close to death.

We therefore suggest that the most efficient way to triage patients for ART should be to give priority to those most likely to die within a specified time, given all of the information that we are able to collect at the time that the person is diagnosed. More precisely we will say that people should be started on ART if their chance of dying in a time $\tau *$ is greater than some pre-determined probability $\pi *$. If we were to choose people at random then, since the mean life expectancy after infection without ART would be about ten years, the average annual mortality would be about 10% and people would, on average, have a 10% probability of dying in the next one year. In order to identify those

2/11

who have a greater than average chance of dying in the next one year we might say that if a person is found to be HIV-positive and has a greater than $\pi* = 5\%$ chance of dying in the next $\tau* = 1$ year then they should be started on ART. The values of $\tau*$ and $\pi*$ can be chosen according to the affordability and availability of the drugs. Any person who is not started on treatment immediately and develops an AIDS related condition should start treatment immediately. Having decided on appropriate values of $\tau*$ and $\pi*$ we can then decide on the $CD4^+$ cell count or the viral load at which to start people on ART.

In deciding on the $CD4^+$ count at which people should start on treatment we need to take into account the fact that $CD4^+$ cell counts are known to vary widely within and among different populations of HIV-negative people,[20,40-47] between men and women,[40,48] for many other reasons,[49-51] and with the stage of the epidemic.[20]

In deciding on the viral load at which people should be started on treatment we need to allow for the fact that set-point viral loads, the value of the viral load at which it reaches a more or less stable value after the acute phase of infection, varies widely within populations of HIV-positive people,[19] and is a good predictor of life expectancy after sero-conversion and without access to ART.[52,53]

Finally, to decide on how frequently people should be tested we need to know the incidence of HIV which will depend on age, gender, the population at risk and the stage of the epidemic.[54]

## Data

In this study we draw on a wide range of data including the age of the person, the time course of the epidemic in the country under consideration and the age and gender specific incidence of infection. We use data on the distribution of survival as a function of viral load, on the distribution of $CD4^+$ cell counts in HIV-negative people and the distribution of the set-point viral load in HIV-positive people. We use data on the relationship between the set-point viral load and survival.

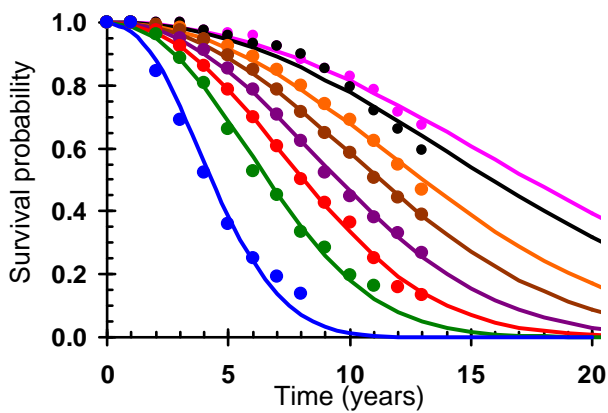

Figure 2. Probability of survival as a function of age at infection.[55] Right (pink) to left (blue) to lines: infected at 0–5, 5–15, 15–25, 25–35, 35–45, 45–55, 55–65, 65+ years. Fits are Weibull probabilities with shape parameter 2.18 ± 0.14. Median survival given in Figure 3.

Survival after infection is highly variable within and between different age groups and the best data are from the CASCADE study[55] shown in Figure 2. The data can be fitted to Weibull survival curves across all age-groups keeping the shape parameter fixed at 2.18 ± 0.14 and allowing the median survival to vary among age-groups. A shape parameter of 2.18 suggests that the survival probability is close to Gaussian and that mortality increases almost linearly with time since infection.

From the fits to the data in Figure 2 we can determine the relationship between the median survival and the age at infection and this is shown in Figure 3. The median survival for a person infected at the age of 30 years is 11.8 years falling linearly from about 16 years for those infected at the age of 5 years, to 5 years for those infected at the age of 65 years.

The distribution of $CD4^+$ cell counts varies both within and among countries. We use data shown in Figure 4 for Botswana,[56] Zimbabwe[46] and South Africa.[20] For Botswana and Zimbabwe we fitted the data to the published median and inter-quartile ranges[57] and for South Africa we used the raw data.[20]

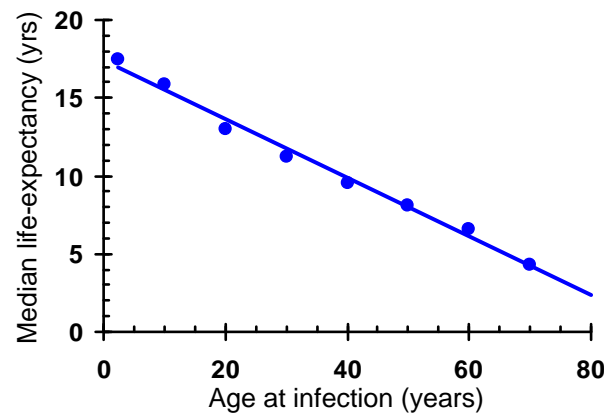

Figure 3. Median life expectancy as a function of the age at infection from the fits to the data in Figure 2. The slope of the line is −0.188 ± 0.008; the value at age 30 years is 11.76 ± 0.18 years.

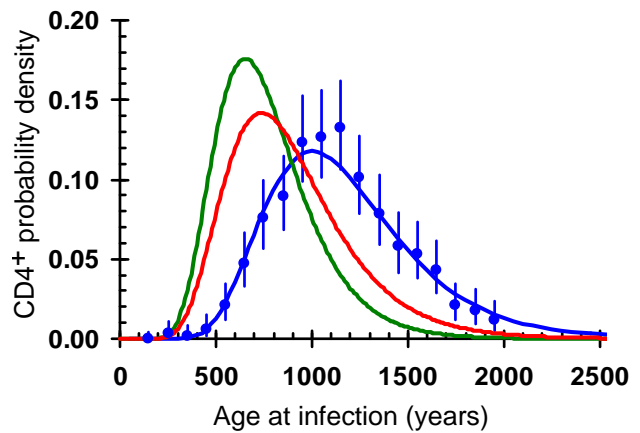

Figure 4. Probability density functions for $CD4^+$ cell counts. Green: Botswana;[56] Red: Zimbabwe;[46] Blue: South Africa.[20] Fitted curves are log-normal distributions with mean $m$ and shape parameter $s$. Botswana $m = 726/\mu L$, $s = 0.33$; Zimbabwe $m = 838/\mu L$, $s = 0.32$; South Africa $m = 1116/\mu L$, $s = 0.33$.

The distribution of viral load in a cross-sectional survey of young men in Orange Farm, South Africa is given in Figure 5[19] and, in the absence of other data we take this to represent the distribution of viral load in all of the countries under consideration. Strictly speaking we should use the distribution of set-point viral loads and the data in Figure 5 will be biased by the shorter life expectancy of people with high set-point viral loads, as illustrated in Figure 6, and also by the time course of the epidemic.

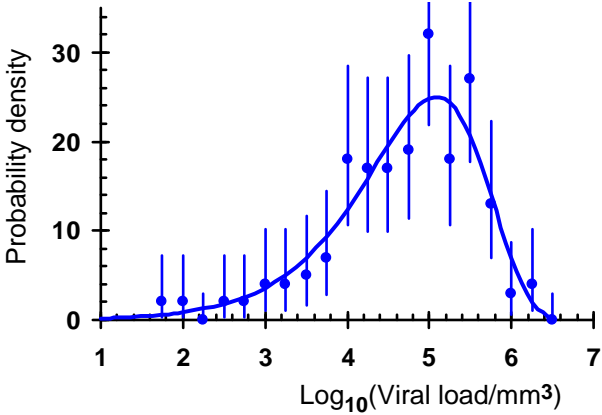

Figure 5. Probability density of viral load as a function of the log-viral load for young men in Orange Farm, South Africa.[19]

The prevalence and incidence of infection over time are shown in Figure 7 and Figure 8, respectively, for Botswana, Zimbabwe and South Africa. These estimates were obtained by fitting a standard SI model, fully described elsewhere, to UNAIDS estimates of the prevalence of HIV over time.[25] The incidence peaked earlier and at a higher level in Zimbabwe than in either Botswana or South Africa but since 1994 has been falling rapidly.[58]

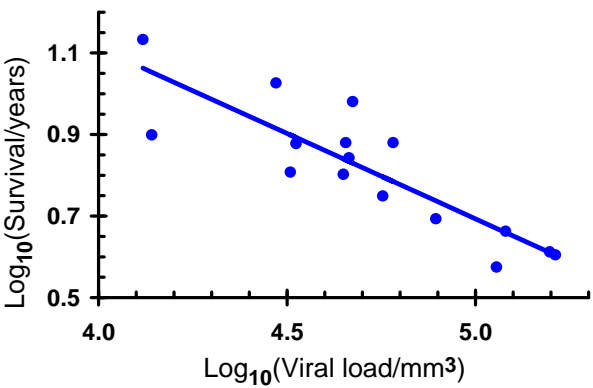

Figure 6. The time from infection to death for 16 individuals plotted as a function of their set-point viral load.[59] The mean value of the survival is $6.52 \pm 0.31$ years and the slope of the line is $-0.419 \pm 0.066$ so that for each ten-fold decline in viral load survival increases by a factor of $10^{0.419} = 2.62 \pm 0.40$. Baeten et al. obtained a slope of $-0.32 \pm 0.25$.[60]

The age-specific prevalence of infection for women attending ante-natal clinics in South Africa in 2010 is shown in Figure 9.[61] Across a number of studies in South Africa it has been shown that distribution of the age-specific prevalence does not vary greatly in different populations and over time[54] and we take this to represent the age-specific prevalence in all three countries at all times.

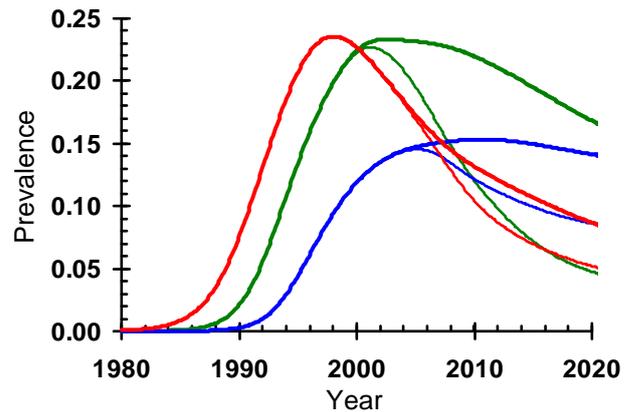

Figure 7. Prevalence of HIV infection in adults aged 15+ years. Red: Zimbabwe; Green: Botswana; Blue: South Africa. Estimates obtained by fitting an SI model, described elsewhere, to UNAIDS estimates of the prevalence of HIV over time.[25] Heavy lines: total prevalence; light lines: prevalence of those not on ART.

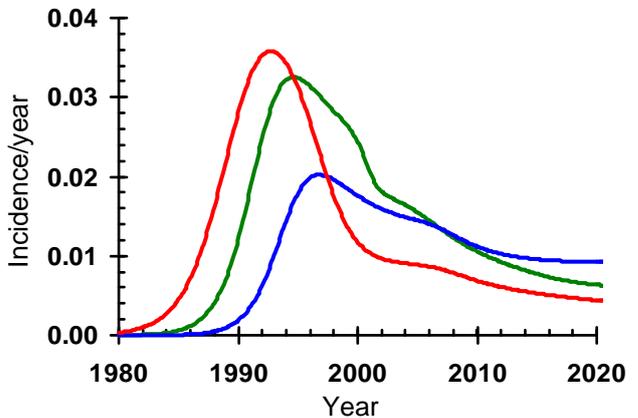

Figure 8. Incidence of HIV infection in adults aged 15+ years. Red: Zimbabwe; Green: Botswana; Blue: South Africa. Estimates obtained by fitting an SI model, described elsewhere, to UNAIDS estimates of the prevalence of HIV over time.[25]

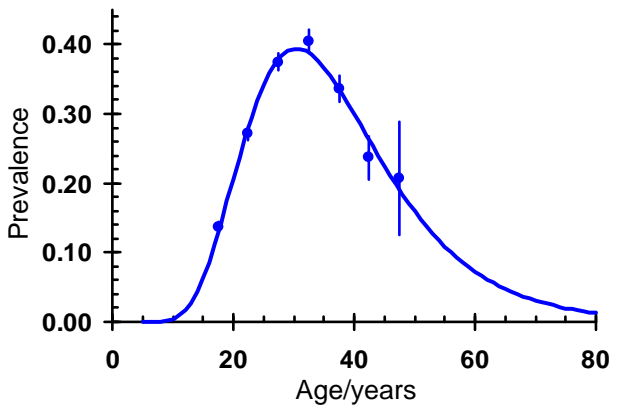

Figure 9. Age specific prevalence of infection for women attending ANC clinics in South Africa in 2010.[61] The fitted curve is a log-normal distribution, off-set by 10 years, a location parameter of 34.9 years and a scale parameter of 0.369.

From the age specific prevalence and the time-trends in the overall prevalence we are able to estimate the age-specific incidence.[62] We use the incidence estimated in this way for men and women in Carletonville, South Africa[54] to represent the age specific incidence in all three countries at all times as shown in Figure 10. We scale this to match the overall incidence shown in Figure 8 for each country.

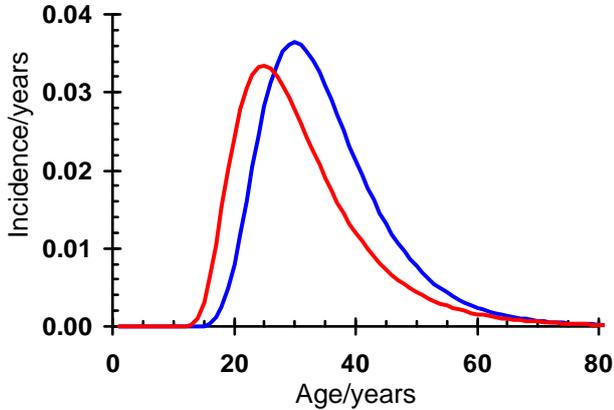

Figure 10. Age specific incidence of infection for women attending ANC clinics in South Africa in 2010. Calculated[62] using data[54] from Carletonville, South Africa.

There are two key distributions in the model: the probability distribution of the time for which a person found to be HIV-positive in a cross-sectional survey has already been infected and the probability distribution of a person's life-expectancy, given the length of time for which they have been infected when they are found to be HIV-positive.

To calculate the probability density of the time since infection, shown in Figure 11, we proceed as follows. Let annual the incidence of HIV at time $t$ and age $a$ be $I(t,a)$, the probability of surviving for $t$ years, given that they were infected at age $a$, be $S(t,a)$,[62] and the probability that they start ART between time $t_1$ and $t_2$ be $A(t_2,t_1)$. Then the probability that a person was infected at time $t-\tau$, is still alive and not on ART, $\overline{\text{ART}}$, at time $t$, which we shall call $R(\tau,t)$, is

$$P\left(\text{Infected at time } t-\tau, \text{alive at time } t, \overline{\text{ART}}\right) = $$
$$I(t-\tau, a-\tau) S(\tau, a-\tau) \overline{A(t, t-\tau)} = \quad 1$$
$$P\left(\text{Infected at time } t-\tau | \text{alive at time } t, \overline{\text{ART}}\right) \times$$
$$P\left(\text{alive at time } t, \overline{\text{ART}}\right)$$

Then since

$$P\left(\text{alive at time } t, \overline{\text{ART}}\right) = $$
$$\int_{-\infty}^{t} P\left(\text{Infected at time } \tau | \text{alive at time } t, \overline{\text{ART}}\right) d\tau \quad 2$$

we can treat this as a normalization factor, $N$. Then

$$R(\tau,t) \equiv P\left(\text{Infected at time } t-\tau | \text{alive at time } t, \overline{\text{ART}}\right) = $$
$$I(t-\tau, a-\tau) S(\tau, a-\tau) \overline{A(t, t-\tau)}/N \quad 3$$

We note firstly that $I(t,a)$ is the incidence per person, not per susceptible person, and secondly that

$$\overline{A(t,t-\tau)} = \exp\left(-\int_{t-\tau}^{t} \frac{I_A(t)}{Q_A(t)} dt\right) = \exp\left(-\int_{t-\tau}^{t} \frac{-\frac{d}{dt} Q_A(t)}{Q_A(t)} dt\right)$$
$$= \exp\left[\ln(Q_A(t)) - \ln(Q_A(t-\tau))\right] \quad 4$$
$$= \frac{Q_A(t)}{Q_A(t-\tau)} = \frac{1-P_A(t)}{1-P_A(t-\tau)}$$

where $I_A(t)$ is the rate at which people start ART, $Q_A(t)$ is the proportion of HIV-positive people who are not on ART and $P_A(t)$ the proportion who are on ART.[†] Figure 11 show the estimate values of $R(\tau,2010)$ for women in South Africa aged 20, 25, 30 45, 60 and 70 years.

In Figure 11 the curve for 20 year-old women is sharply peaked at the origin. Because women only become sexually active at a median age of 16 years,[63] any 20 year old women who are HIV-positive must have been recently infected. The curve for 70 year old women is also concentrated near the origin because older people have a very short life-expectancy after infection (Figure 3) so they too are more likely to have been infected recently. It is important to remember that while 25 year-old women and 70 year old-women have a very similar distribution of the time since they were infected, their life expectancies when found to be HIV-positive are very different.

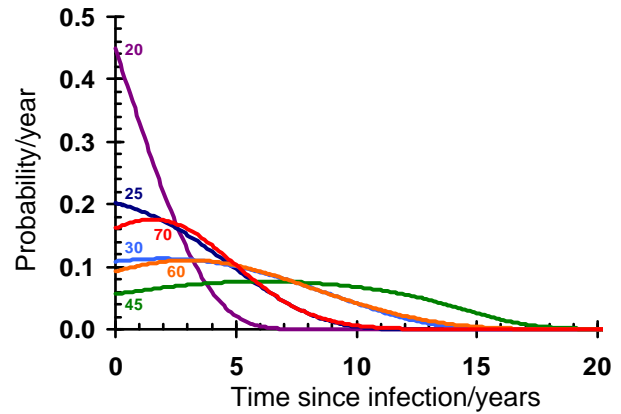

Figure 11. Probability density of time since infection, $R(t)$, for women in South Africa, not on ART, aged 20, 25, 30, 45, 60 and 70 years of age in 2010. Coloured numbers in the chart give the age at which the person is tested for HIV for the corresponding line.

## Model

The model is constructed as follows. We assume that a person of a given age and gender is tested for HIV in a given country and in a given year. We first calculate the probability density of the time since they were infected as descried in the data section and illustrated in Figure 11. This depends on the incidence of infection over time since they were more likely to be infected when the incidence was high than when it was low (Figure 8); on the age-specific incidence of infection since they were more likely to be infected between the ages of 25 and 30 years than at younger or older ages (Figure 10); on their survival as a function of their age at infection since older people are more likely to have died before being tested (Figure 3); and on the proportion of HIV-positive people who are on ART (Figure 7) since if they have already started ART they do not need to start again. Having calculated the probability density of the time since they were infected, we choose a random value, $R$, from this distribution.

---
[†] This approximation does not account for the number of people that become newly infected with HIV or that die before starting ART.

We then use the known distribution of survival as a function of their age at infection (Figure 2 and Figure 3) and we choose a random value, $S$, from this distribution.

Finally we obtain a value for their $CD4^+$ cell count when they are tested. To do this we choose a random value from the $CD4^+$ cell count distribution of people in that country, $C$, reduce it by 25% to allow for the impact of the acute phase,[20] and assume that after the acute phase their $CD4^+$ cell counts decline linearly to death if they are not treated. From the values of $R$, $S$ and $C$, it is a trivial matter to calculate the corresponding value of the $CD4^+$ cell count when they are tested for HIV.

We then repeat the process very many times, typically $10^5$ to $10^6$ times, to determine the probability density of their life-expectancy, if they are not treated, as a function of their $CD4^+$ cell count when they are tested.

Finally, we calculate the $CD4^+$ cell count at which the probability of dying in the next $\pi*$ years is greater than $\pi*$ and this gives the optimal $CD4^+$ cell count at which to start treatment for people of a given age in a given country in a given year.

We carry out essentially the same exercise assuming that we know their viral load rather than their $CD4^+$ cell count. To do this we need a model for survival as a function of viral load but there is an important, unresolved question about which we have to make an assumption. We know that viral load and age are both good predictors of survival after infection with HIV (Figure 3 and Figure 6). What is not known is whether or not these factors are independent or correlated. In the absence of better data we have two choices: a) we can assume that the set-point viral load increases with age and that this explains the dependence of survival on the age at infection, or b) we can assume that the set-point viral load does not change with age at infection so that survival depends on both set-point viral load and age at infection. Data on survival as a function of viral load are sparse and such studies can no longer be repeated given the availability of ART. However, it should be possible to obtain data on the distribution of the set-point viral load as a function of the age at infection and this might clarify the situation.

In this study we follow the first of the two assumptions in the previous paragraph and scale the initial distribution of viral load (Figure 5) so that the expected survival for the mean value (Figure 6) is equal to the expected survival for that age (Figure 3). The alternative, the second of the two assumptions in the previous paragraph, which we have not explored here would be to assume that the set-point viral load is independent of age at infection and then allow the distribution of survival as a function of viral load to shift to shorter survival times as the age at infection increases.

Using the model we also calculate the proportion of people that would have died within one year if they had not started ART, the coverage of ART, that is to say the proportion of people in each age group that would need to start ART under the conditions specified, and the reduction in transmission as a result of people starting ART. We note that the reduction in transmission is calculated as the time between starting treatment and death divided by the time between infection and death if treatment had not been provided. This does not allow for the fact that people with high viral loads are more infectious than those with low viral loads.[64]

Finally, we consider the frequency with which people should be retested. To do this we calculate the time it would take for them to be infected with HIV with a probability of 5% using the current incidence of HIV (Figure 8), the age specific incidence of HIV Figure 10) and their age. Note that we assume that the incidence of infection is constant rather than decreasing over time so that these estimates will be lower bounds.

## Results

We first consider what would have been the most appropriate $CD4^+$ cell count at which to start ART if the drugs had been available in 1990 as shown in Figure 12 (top graph). In the youngest age group those in Botswana should have started ART if their $CD4^+$ cell count was below 350/μL while those in South Africa should have started when their $CD4^+$ cell count was below 550/μL. The $CD4^+$ cell count at which people should have started treatment increases slowly up to the age of about 40 years (South Africa) and 50 years (Botswana) but all those aged above 50 years should have started treatment immediately.

The key point to note is that in all three countries, but especially South Africa, the epidemic was relatively young in 1990 so that young people, who had only recently become sexually active, must have been recently infected. The difference between the $CD4^+$ cut-off in Botswana and South Africa reflects the substantial difference in the $CD4^+$ cell counts in the two countries. Since life-expectancy, after infection falls with age, older people need to start earlier, and therefore at higher $CD4^+$ cell counts, than younger people.

If we now consider the situation in the year 2000 as shown in Figure 12, middle graph, the criterion for the youngest age group remains unchanged since they had to have been very recently infected. Older people, who have been sexually active for longer, are more likely to have been infected for longer (Figure 8), and are therefore more likely to die sooner. In all three countries everyone above the age of 30 to 35 years should have started ART immediately on diagnosis. The situation in 2010 is similar to that in 2000 but with the age at which everyone should have been started on ART shifted to younger ages.

The proportion of people who would have died within one year without treatment (brown text above each graph in Figure 12) is, by definition, 5% if the $CD4^+$ criterion is used but once immediate treatment is recommend the proportion that would have died without treatment increases with age to between 10% and 20% at age 70 years. The coverage, that is to say the proportion of people who test positive and need to start treatment at each age (violet text above each graph in Figure 12) increases from about 30% in the youngest age group to 100% in the oldest.

The reduction in transmission, that is to say the proportion by which their infectious period is reduced because they are put onto ART (orange text above each graph in Figure 12), increases from about 15% in the youngest age group to between 50% and 70% in the oldest age groups.

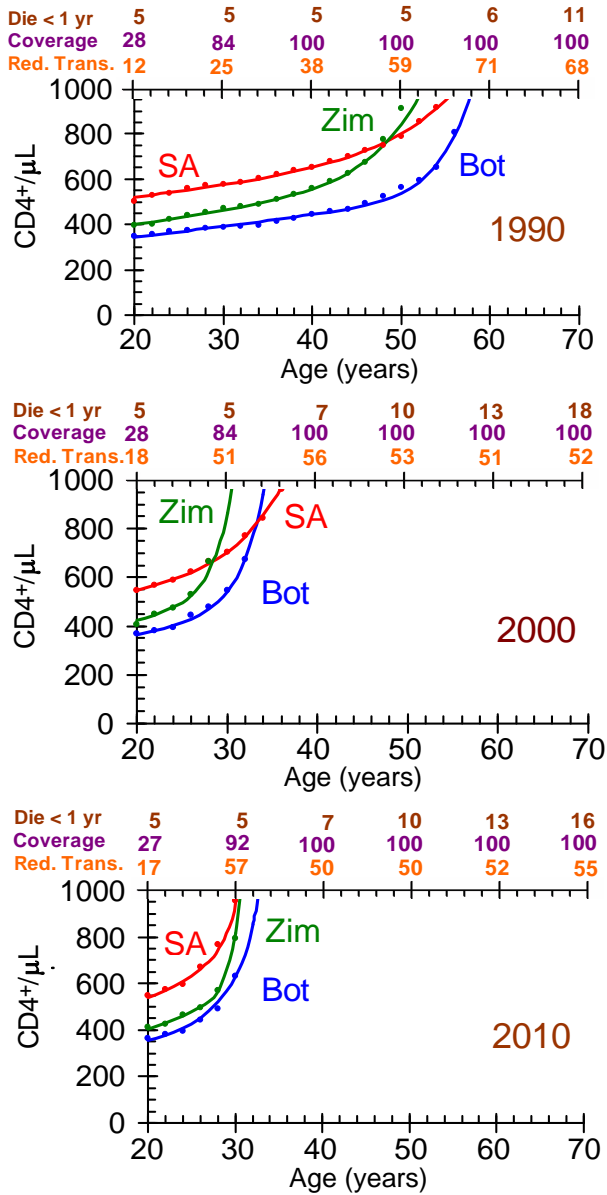

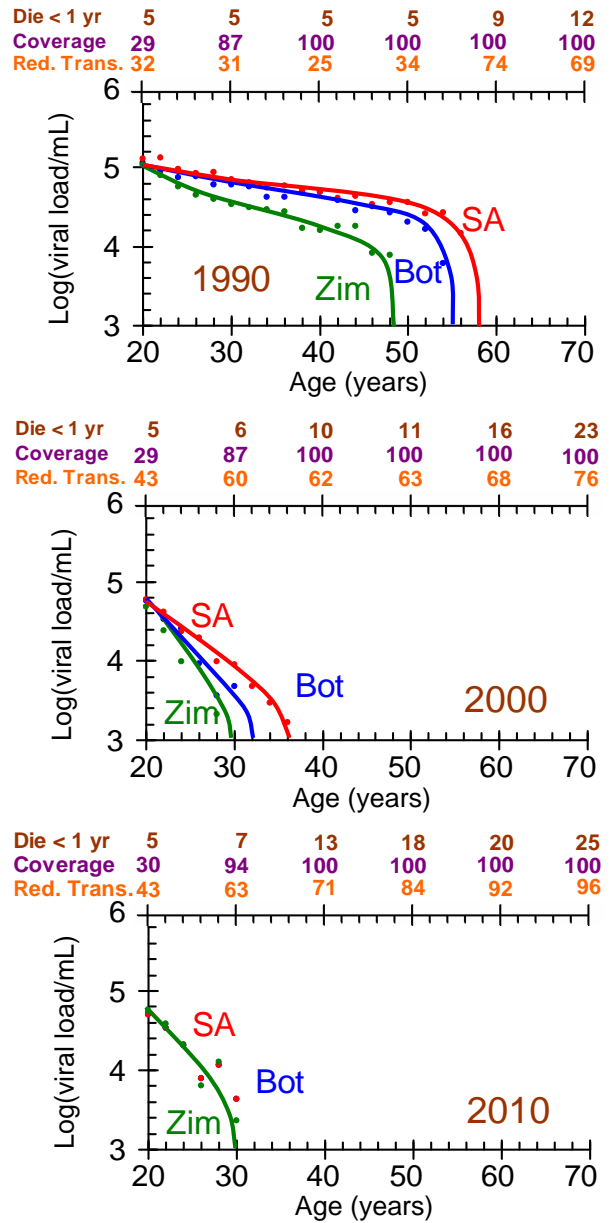

Figure 12. The $CD4^+$ cell count at which people who are tested randomly and found to be HIV-positive should start ART, as a function of their age, if they are to have a 95% chance of surviving for more than one year without it. From top to bottom: testing done in 1990, 2000 and 2010. Data shown for Botswana, South Africa and Botswana. The numbers above each graph are: Brown: the proportion that will be expected to die within one year, without treatment, if people only start ART at the indicated $CD4^+$ cell count; Violet: the proportion of all those testing positive who would be started on ART; Red: the amount by which transmission would be reduced if people only start ART at the indicated $CD4^+$ cell count. Numbers above graphs are averages over all three countries.

Figure 13. The $CD4^+$ cell count at which people who are tested randomly and found to be HIV-positive should start ART, as a function of their age, if they are to have a 95% chance of surviving for more than one year without it. From top to bottom: testing done in 1990, 2000 and 2010. Data shown for Botswana, South Africa and Botswana. The numbers above each graph are: Brown: the proportion that will be expected to die within one year, without treatment, if people only start ART at the indicated $CD4^+$ cell count; Violet: the proportion of all those testing positive who would be started on ART; Red: the amount by which transmission would be reduced if people only start ART at the indicated $CD4^+$ cell count. Numbers above graphs are averages over all three countries.

The proportion of people who would have died within one year without treatment (brown text above each graph in Figure 13) is, again by definition, 5% if the CD4+ criterion is used but once immediate treatment is recommend the proportion that would have died without treatment increases with age to between 10% and 20% at age 70 years.

The coverage, that is to say the proportion of people who test positive and need to start treatment at each age (violet text above each graph in Figure 12) increases from about 30% in the youngest age group to 100% in the oldest.

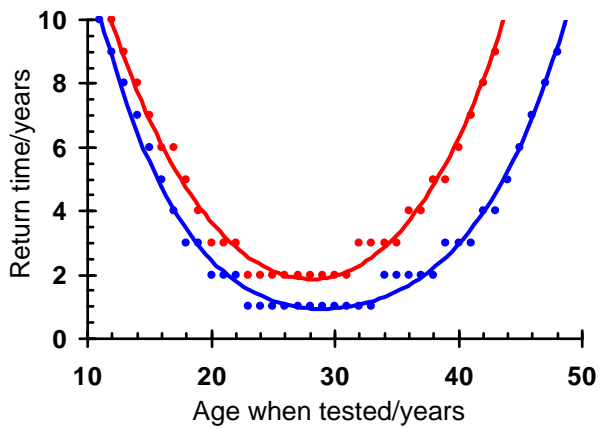

Figure 14. The time at which a person should be asked to return for testing if they are found to be HIV-negative as a function of age in South Africa and Zimbabwe.

The reduction in transmission, calculated as the proportion by which their infectious period is reduced because they are put onto ART (orange text above each graph in Figure 12), is between about 30% and 40% in the youngest age group but increases to between 70% and almost 100% in the oldest age groups.

Figure 13 shows how soon a person should be asked to return for testing as a function of their age when tested if they are found to be HIV-negative. In South Africa people between the ages of about 20 and 38 years should be asked to come back for testing after one year; people between the ages of 16 to 20 years or 38 to 42 years should be asked to come back for testing after two years; people below that age of 16 or above the age of 42 should be asked to come back after five years. In Zimbabwe the return times could be roughly twice as long.

## Conclusions

In the early days of the HIV epidemic in southern Africa one could have argued that using a CD4+ threshold or a viral load threshold for deciding on when to start treatment was reasonable both for the individual person and as a way of triaging patients in favour of those most urgently in need of treatment. If one had been using a threshold based on CD4+ counts young adults should have been started on ART if their viral load was below about 400/μL in Botswana and 600/μL in South Africa. If one had been using a threshold based on viral load, young adults should have started treatment if their viral load was above 100k/mL in all countries. The early IAS criteria for deciding when to start treatment were about right for young people.

The epidemics of HIV are now mature and the situation has changed. For people aged 20 years one could consider delaying treatment using the same criteria as outlined early in the epidemic. However, it is now in the best interests of the patients to start ART immediately if they are more than 30 years old and found to be HIV-positive. Under these circumstances it would make most practical sense to start all HIV-positive people on treatment immediately.[65] The recommendations made by the World Health Organization between 2002 and 2009 (Figure 1) greatly reduced the number of people who were eligible for treatment.

Although they are not included in this study, a group of people who are of particular concern are those children who are infected vertically and survive for more than two years after which their life-expectancy is about 16 years.[65] However, it is clear that if they are not started on treatment immediately they are unlikely to develop normally, either physically or mentally, and they must be started on ART as early as possible.

It is important to decide on the frequency with which people should be retested especially in settings where the prevalence is low. For example, based on data from women attending ante-natal clinics, the prevalence of infection among low risk women in Vietnam is about 0.3% and for every 1,000 women tested only 3 would be found to have HIV. For such women active case-finding would be expensive but passive case finding and provider initiated testing would be appropriate. Since the prevalence of infection among such women must be about 0.03% per year one might argue that if they are found to be HIV-negative their chance of becoming infected in the next ten twenty years, say, would still be less than 1% and they need not be retested unless of course they have an infected partner or reason to believe that they may have been exposed to HIV.

This analysis suggests that given the current state of HIV in southern Africa it is in the best interests of the individual person to start treatment as soon as they are found to be HIV positive and that neither CD4+ cell counts nor viral load tests will significantly help to decide on whom to treat. The added benefits to the individual of avoiding the deterioration of their immune system which starts as soon as they have become infected as well as the very substantial reduction in the likelihood that they will infect their sexual partners[22,66,67] provides even greater urgency to the need to start treatment as soon as people are found to be HIV-positive.

The most important unknown in the model to assess the impact of using viral load tests for deciding when to start treatment concerns the relationship between survival and viral load. It is thought that both are good predictors of life-expectancy after infection but there are few data that help us to understand the relationship between set-point viral load and age as predictors of survival.

We show in this paper that if we are concerned to 'do no harm' then the use of CD4+ cell counts or viral load testing should be abandoned. If we also take into account the now accepted fact that the HIV-virus begins to compromise ones immune system from within weeks of infection, that the immune function, even with the best available drugs, shows significantly less recovery if one starts treatment later,[23] the increase in mortality even among those with very high $CD4^+$ cell counts, the very substantial evidence that early treatment stops transmission,[19,22,68] and the cost and social savings that will undoubtedly accrue from early treatment,[34,36] then regular testing and early treatment should now be made the standard of care.

## Acknowledgments

We thank Eleanor Gouws for her advice and comments on this manuscript.